# Core Noise Requirements, and GW Sensitivities of AMIGO

Wei-Tou Ni

*National Astronomical Observatories, Chinese Academy of Sciences, Beijing, China*
*Innovation Academy of Precision Measurement Science and Technology (APM),*
*Chinese Academy of Sciences, Wuhan 430071, China*

AMIGO – The Astrodynamical Middle-frequency Interferometric GW (Gravitation-Wave) Observatory is a first-generation mid-frequency GW mission bridging the sensitivity gap between the high-frequency GW detectors and low-frequency space GW detectors. In our previous works, we have obtained appropriate heliocentric orbit formations of nominal arm length 10,000 km with their first-generation time-delay configurations satisfying frequency noise reduction requirement, and we have also worked out thrust-fuel friendly constant-arm heliocentric orbit formations. In this paper, we review and study noise requirements and present the corresponding GW sensitivities. From the design white position noises and acceleration noises, we obtain the GW sensitivities for the first-generation Michelson X TDI configuration of b-AMIGO (baseline AMIGO), AMIGO, and e-AMIGO (enhanced AMIGO). In view of the current technology development, we study and indicate steps to implement the AMIGO mission concept.

# 1. Introduction

The mid-frequency GW (Gravitation-Wave) band (0.1-10 Hz) [1] between LIGO-Virgo-KAGRA detection band and LISA-TAIJI-TIANQIN detection band is rich in GW sources [2-5]. In addition to the intermediate BH (Black Hole) binary coalescences (The BH binary coalescence event GW190521 with a total mass of 150 $M_{sun}$ has been detected by LIGO-Virgo collaboration recently [6].), the inspiral phases of stellar-mass coalescences and GWs from compact binaries falling into intermediate BHs, it also enables us to study the compact object population, to test general relativity and beyond-the Standard-Model theories, to explore the stochastic GW background, etc. [2-6] In addition to DECIGO and BBO, the detection proposals under current study for mid-frequency GW include AEDGE, AIGSO, AION, AMIGO, DO, ELGAR, INO, MAGIS, MIGA, SOGRO, TIANGO, TOBA, ZAIGA. (See, e.g., [5] for a brief review). The strain ASDs (amplitude spectral densities) versus frequency for various GW detectors and GW sources are plotted in Figure 1.

In our previous work [7, 20, 21], we have worked on mission concept and orbit design of AMIGO of arm length $10^7$ m. Depending on the stringency of the noise requirements, there are three versions of AMIGO – baseline AMIGO (b-AMIGO), AMIGO, and enhanced AMIGO (e-AMIGO). In this paper, we focus on the study of noise requirements and present the sensitivity curves. Section 2 introduces these requirements, and presents GW sensitivities based on the requirements. Subsection 2.1 lists three heliocentric AMIGO orbit choices obtained in [7]. Subsection 2.2 lists core noise requirements for b-AMIGO, AMIGO and e-AMIGO. In Section 3 we study preliminarily the technology outlook and how to reach these requirements. With respect to the current technology development, we study and indicate steps to implement the AMIGO mission concept. We also present the noise requirements for the sensitivities of AMIGOs with arm length $5 \times 10^7$ m and discuss the trade-offs among AMIGOs with different arm lengths.

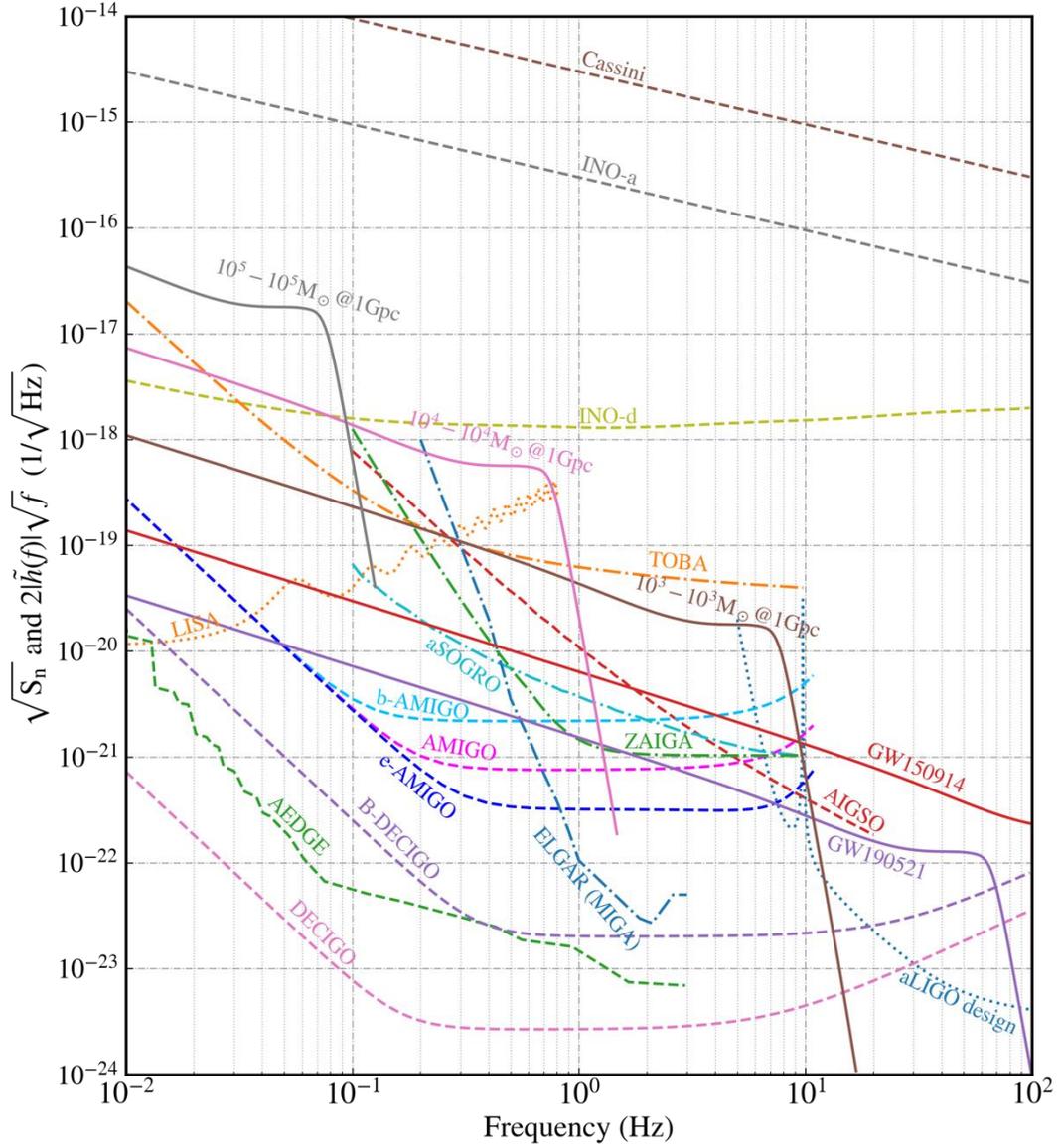

**Fig. 1.** Strain ASDs (amplitude spectral densities) vs. frequency for various GW detectors and GW sources. The solid lines show the inspiral, coalescence and oscillation phases of GW emission from various equal-mass black-hole binary mergers in circular orbits at a distance of 1 Gpc. The strain ASD of GW150914 is calculated from its characteristic amplitude in Figure 1 of [8] using standard formula. The strain ASD of GW190521 is calculated from the parameters obtained in [6]. The AMIGO design sensitivity is in dashed red while basic AMIGO (b-AMIGO) and enhanced AMIGO (e-AMIGO) sensitivities are in dashed green and dashed blue, respectively. The three curves merge at lower frequency in the figure; the label AMIGO in Figure 1 of [7] is a mis-label and should be revised to b-AMIGO. (AEDGE: [9]; CSDT: Cassini Spacecraft Doppler Tracking [10, 11]; INO: Interplanetary Network of Optical lattice clocks [12]; AIGSO: Atom-Interferometric GW Space Observatory [13, 14]; TOBA: Torsion Bar Antenna [15, 16]; MIGA/ELGAR: Matter-Wave laser Interferometer Gravitation Antenna [17] / a European Laboratory for Gravitation and Atom-interferometric Research [18]; ZAIGA: Zhaoshan long-baseline Atom Interferometer Gravitation Antenna [19]).   Credit: this figure is updated by Gang Wang from Figure 1 of [7] .

## 2. AMIGO

In Ref. [7], we have worked out three heliocentric orbit configurations and two geocentric orbit configurations for AMIGO. In [7], we have also addressed the issue whether the orbit design of constant-arm versions of AMIGO are feasible, and have obtained the induced acceleration and thruster requirements. For the solar-orbit options, the acceleration to maintain the formation can be designed (i) to be less than 15 nm/s$^2$ with the thruster requirement in the 15 μN range for the first orbit option listed in Subsection 2.1; (ii) to be less than 50 nm/s$^2$ with the thruster requirement in the 50 μN range for the second orbit option listed in Subsection 2.1; (iii) to be less than 500 nm/s$^2$ with the thruster requirement in the 500 μN range for the third orbit option listed in Subsection 2.1. The propellent requirements are well manageable. For the 2 geocentric orbit options, the required accelerations are more than 3 orders larger than 3 heliocentric options with requirements on the propellent mass not deliverable. The gravity gradients in the geocentric environments are much higher than the heliocentric environments. We choose heliocentric options in this study. For heliocentric choices, AMIGO would be a good place to test the feasibility of the constant equal-arm GW detection in addition to the first generation TDI configuration GW detection. The test mass acceleration actuation requirement will be briefly considered in relation to acceleration noise requirement in Section 3. *From the orbit study, the solar orbit option is the mission orbit preference. Three choices of heliocentric options are listed in Subsection 2.1.*

We have studied the deployment for heliocentric orbit options. There are two desirable options for the *third orbit case in Subsection 2.1*: (i) A last-stage launch from 300 km LEO (Low Earth Orbit) to an appropriate 2-degree-behind-the-Earth AMIGO formation [(iii) in Subsection 2.1] in 95 days with a Δv of about 80 m/s when arrived to reach the science orbit velocity for the 3 S/C; (ii) A last-stage launch from 300 km LEO (Low Earth Orbit) to an eccentric Hohmann orbit with apogee 262931 km (The period of this Hohmann transfer orbit is about 6 days.) (Fig. 2). It takes 3 days (half an elliptic orbit) for this transfer from perigee to apogee (Fig. 2. (b)). This apogee is designed to be the closest encounter point with respect to Earth of the center of mass of the 3 S/C traced back in time geodetically of the 2-degree-behind-the-Earth AMIGO formation [(iii) in Subsection 2.1]. From here, a Δv's of about 1.6 km/s are needed to allow the 3 S/C to enter their respective (pre-)science orbits (Fig. 2. (c)). *It takes less than a week from launch to reaching the science orbit.* After entering the (pre)-science orbits around the 262,931 km apogee, the calibration, commissioning and various science operations can be started [7, 22]. As to the deployment for the first and second orbit cases in Subsection 2.1, we don't have the second choice above. However, we do have the first choice. Ref. [22] will present the details; see also Ref.'s [23, 24] for similar deployments in the cases for LISA and ASTROD-GW.

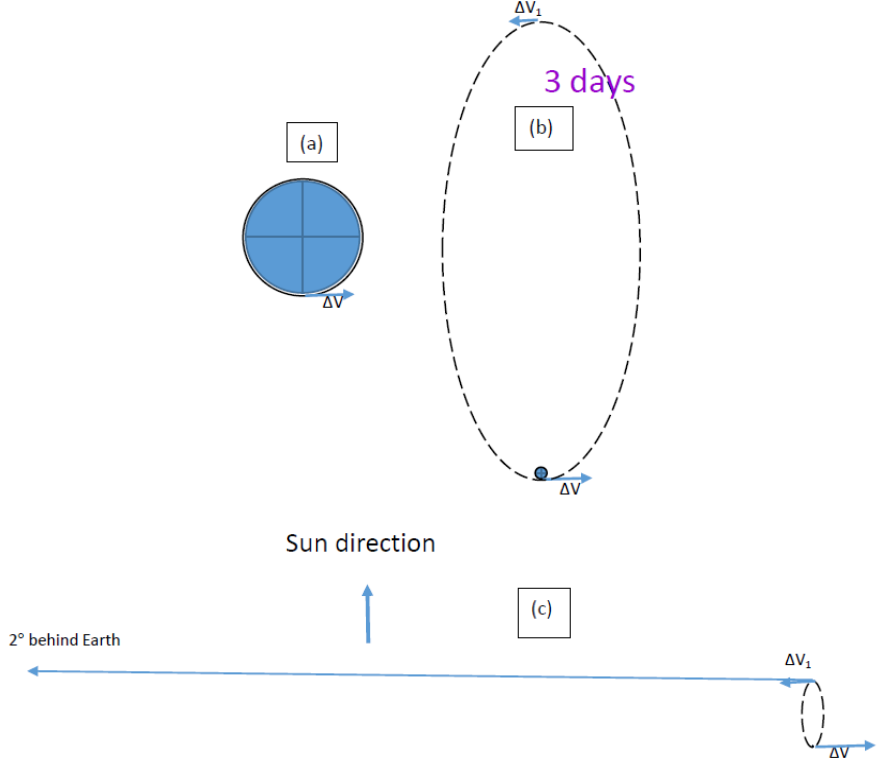

**Fig. 2.** An option for taking less than a week from launch to reaching the (pre-)science orbit for AMIGO-S-2-4deg. (a) Launch to 300 km LEO (Low Earth Orbit) parking orbit. (b) A last-stage launch from 300 km LEO to an eccentric Hohmann orbit with apogee 262931 km (The period of this Hohmann transfer orbit is about 6 days.); It takes 3 days (half an elliptic orbit) for this transfer from perigee to apogee. (c) This apogee is designed to be the closest encounter point with respect to Earth of the center of mass of the 3 S/C traced back in time geodetically of the 2-degree-behind-the-Earth AMIGO formation [(iii) in Subsection 2.1]. From here, a $\Delta v_1$ of about 1.6 km/s are needed to allow the 3 S/C to enter their respective (pre-)science orbits (Fig. 2. (c)).

For the laser-interferometric space GW detection, the acceleration/inertial-sensing noises and the laser metrology noises are the core (key) noises when the contribution to laser-frequency noises are below the core noise levels due to the close match of the two interfering paths. Either the equal-arm length Michelson or the time-delay interferometry (TDI) could possibly do that. LISA adopts the TDI method, while AMIGO is open to both equal-arm length method and TDI method. With the success of launch and first stage of science operation of LISA Pathfinder [25], LISA of new arm length of 2.5 Gm has been proposed by LISA Consortium [26] and approved by ESA. At the time, LISA acceleration/inertial-sensing noise goal was set below the colored acceleration noise level over the frequency range 20 μHz $< f <$ 1 Hz [27, 28]:

$$S_a(f) = 9 \times 10^{-30} [1 + (10^{-4}\ \text{Hz}/f)^2 + 16\ (2 \times 10^{-5}\ \text{Hz}/f)^{10}]\ \text{m}^2\ \text{s}^{-4}\ \text{Hz}^{-1}. \qquad (1)$$

In our original 2017 proposal for AMIGO [20], we adopted (1) as achievable goal for AMIGO acceleration/inertial-sensing noise level. In our interested frequency range 10 mHz < $f$ < 10 Hz of AMIGO, the formula (1) is different from the non-colored constant formula

$$S_a(f) = 9 \times 10^{-30} \text{ m}^2 \text{ s}^{-4} \text{Hz}^{-1/2}, \quad \text{for } 10 \text{ mHz} < f < 10 \text{ Hz}, \qquad (2)$$

by only ~0.01 %. With further measurements of LISA Pathfinder, the lower frequency part of the LISA acceleration/inertial-sensing noise requirement is fully satisfied for $f$ < 30 mHz [29]. Above 30 mHz, the experimental acceleration/inertial-sensing noise of LISA Pathfinder is limited by the laser-interferometric readout noise (laser metrology noise) which translates into an effective acceleration/inertial-sensing noise proportional to $(34.8 \pm 0.3)$ fm × $(2\pi f)^2$ Hz$^{-1/2}$ [25, 29]. A blue-coloured factor $[1 + (f/f_c)^4]^{1/2}$ with $f_c$ about 8 mHz is included to relax the original requirement in the new LISA requirement [30]. Since LISA sensitivity is limited by antenna response on the high frequency part, inclusion of this blue-coloured factor does not affect the sensitivity curve significantly. To improve the upper frequency part of the acceleration/inertial-sensing noise spectrum for AMIGO, requirement on the laser metrology noise needs to be more stringent than LISA. We set the AMIGO blue-coloured factor of $S_a(f)$ to be $[1 + (f/0.3 \text{ Hz})^4]^{1/2}$, i.e.

$$S_a(f) \leq 9 \times 10^{-30} [1 + (f/0.3 \text{ Hz})^4] \text{ m}^2 \text{ s}^{-2} \text{ Hz}^{-1/2}, \quad (10 \text{ Hz} > f > 10 \text{ mHz}), \qquad (3)$$

and will discuss more on this in Section 3.

The present noise requirement on the arm laser metrology on AMIGO with $10^7$ m arm length are the same as listed in [7] for b-AMIGO, AMIGO and e-AMIGO. For AMIGO with $5 \times 10^7$ m arm length to be discussed in Section 3 on the optimal arm length to bridge the sensitivity gap between LIGO-Virgo-KAGRA detection band and LISA-TAIJI-TIANQIN detection band, we relax the arm laser metrology requirement by the ratio of the arm lengths (to keep the same emitting laser power and telescope diameter).

With frequency noise and other noises contributed/suppressed below the core noise levels, the GW sensitivity can be well given by the following formula for most of purposes:

$$S_{\text{AMIGOn}}^{1/2}(f) = (20/3)^{1/2}(1/L_{\text{AMIGO}}) \times [(1+(f/(1.29 f_{\text{AMIGO}}))^2)]^{1/2} \times [(S_{\text{AMIGOp}}+4S_a/(2\pi f)^4)]^{1/2}, \qquad (4)$$

where $L_{\text{AMIGO}}$ is the arm length and $f_{\text{AMIGO}} = c/(2\pi L_{\text{AMIGO}})$ is the critical (characteristic) frequency of the detector. Equation (4) is a good approximation of an X TDL-configuration interferometry of a triangular formation for GW sensitivity averaged over sky position and polarization. For two independent X TDL-configuration as in a triangular formation, the strain sensitivity is enhanced by a factor of $(2)^{1/2}$ and the factor $(20/3)^{1/2}$ becomes $(10/3)^{1/2}$ as sometimes quoted. See Ref. [30] and references therein for a derivation. It is also a good approximation for a classical Michelson interferometry [31-34] of a triangular formation. It is the formula we used in our previous papers [7, 20].

## 2.1. Orbit

Three heliocentric AMIGO orbit choices in [7] are listed below:

(i) AMIGO-S-8-12deg: AMIGO-Earth-like solar orbits with formation varying between 8 and 12 degrees behind the Earth orbit starting at epoch JD2462316.0 (2029-Jun-28th 12:00:00),

(ii) AMIGO-S-2-6deg: AMIGO-Earth-like solar orbits with formation varying between 2 and 6 degrees behind the Earth orbit starting at epoch JD2462416.0 (2029-Oct-6th 12:00:00),

(iii) AMIGO-S-2-4deg: AMIGO-Earth-like solar orbits with formation varying between 2 and 4 degrees behind the Earth orbit starting at epoch JD2462503.0 (2030-Jan-1st 12:00:00), in J2000 mean-equatorial solar-system-barycentric coordinate system.

## 2.2. Noise requirement

In this subsection, we set the noise requirement for AMIGO. For the acceleration noise requirement, we set

$$S_a^{1/2}(f) \leq 3 \times 10^{-15} [1 + (f/0.3 \text{ Hz})^4]^{1/2} \text{ m s}^{-2} \text{ Hz}^{-1/2}, \quad (10 \text{ Hz} > f > 10 \text{ mHz}), \tag{5}$$

N.B. LISA LPF has already demonstrated in Feb 2017,
$$S_a(f) \leq 9 \times 10^{-30} [1 + (10^{-4} \text{ Hz}/f)^2 + 16 (2 \times 10^{-5} \text{ Hz}/f)^{10}] \text{ m}^2 \text{ s}^{-4} \text{ Hz}^{-1}, (20 \text{ μHz} - 0.03 \text{ Hz}) \tag{6}$$

For laser metrology noise, we set:

Baseline (b-AMIGO): $S_{\text{AMIGOp}} \leq 1.4 \times 10^{-28} \text{ m}^2 \text{ Hz}^{-1}$, \hfill (7)
$S_{\text{AMIGOp}}^{1/2} \leq 12 \text{ fm Hz}^{-1/2}$, (10 Hz > $f$ > 10 mHz)

Design Goal (AMIGO): $S_{\text{AMIGOp}} \leq 0.14 \times 10^{-28} \text{ m}^2 \text{ Hz}^{-1}$, \hfill (8)
$S_{\text{AMIGOp}}^{1/2} \leq 3.8 \text{ fm Hz}^{-1/2}$, (10 Hz > $f$ > 10 mHz)

Enhanced Goal (e-AMIGO): $S_{\text{AMIGOp}} \leq 0.0025 \times 10^{-28} \text{ m}^2 \text{ Hz}^{-1}$, \hfill (9)
$S_{\text{AMIGOp}}^{1/2} \leq 0.5 \text{ fm Hz}^{-1/2}$, (10 Hz > $f$ > 10 mHz)

## 3. Discussion and Outlook

In this section, we discuss and indicate steps to implement the AMIGO mission requirements. First, we review briefly the current technology achieved for the optical path metrology.

(i) After 3-month operation time since the start of scientific operations on March 1 2016, the LISA Pathfinder Team reported that the measured interferometer displacement readout noise above 60 mHz to 5 Hz is 35 fm Hz$^{-1/2}$, more than two-order below the required 9 pm Hz$^{-1/2}$ [25, 35, 36].

(ii) In the prerecorded talk on Sensor noise in LISA Pathfinder of 2020 LISA Symposium, Gudrun Wanner showed a slide on *the differential TM (Test Mass) displacement noise: June 1st 2016*. In the frequency range between 0.2 Hz to 5 Hz, the measured displacement noise is 31.9±1.7 fm Hz$^{-1/2}$ agreed to OMS (Optical Metrology System) model [36]. The OMS model contains shot noise, relative intensity noise, frequency noise, TM Readout noise, and thermally driven noise. Below 0.2 Hz, the excess noise includes TM Brownian motion, and TM alignment (TTL [tilt-to-length] coupling) noise.

(iii) TTL coupling in an important noise source of space interferometry [37-39]. In LISA Pathfinder's first measurements, a bulge in the acceleration noise appeared in the 20-200 mHz frequency region. This bulge was due to S/C motion coupling into the longitudinal readout. Wanner *et al.* [37] showed that this TM alignment (TTL coupling) noise could be subtracted out.

(iv) GRACE (Gravity Recovery and Climate Experiment) [40] with two satellites, one trailing the other about 200 km apart in a near-polar orbit at approximately 500 km altitude, measured Earth's time-variable gravity field successfully from 2002 until 2017. GRACE Follow-On (GRACE-FO), the successor mission [41] designed as an almost identical copy of GRACE, launched in May 2018. Both missions use a K/Ka band microwave ranging to measure the distance variations between the two satellites for determining the Earth's time-variable gravity field. In addition, GRACE-FO has a Laser Ranging Interferometer (LRI) to measure the same observable, but with higher accuracy, and serves as a technology demonstrator for future geodesy missions and space GW detection missions like LISA [42]. Wegener et al. [38] estimated the TTL couplings of LRI in terms of coupling factors; they are all within 200 μm/rad and meet the requirements.

(v) Chwalla, Danzmann, Alvarez et al. [39] have made a lab demonstration of reduction of TTL coupling by introducing two- and four-lens imaging systems. TTL coupling factors are below ±25 μm/rad (i.e., ±1 pm/40 nrad) for beam tilts within ±300 μrad of the system. They have compensated the additional TTL coupling due to lateral-alignment errors of the imaging system by introducing lateral shifts of the detector. These results help validate the noise-reduction technique for the LISA or other long-arm interferometer. For AMIGO, the TTL coupling should be kept smaller by one more order of magnitude to below ±2.5 μm/rad and the alignment noise should also be kept smaller by one order to within ±30 μrad of the system.

(vi) For the arm metrology, both LISA and AMIGO, in their respective best performance frequencies, require that the noise is basically limited by shot noise. This means that the fractional arm length measurement noise ($\delta L/L$) is inversely proportional to the square root of received power divided by arm length, i.e. inversely proportional to $D_eD_r$ with $D_e$ the diameter of the emitting telescope, $D_r$ the diameter of the receiving telescope and $L$ the arm length (Power received is proportional to $D_e^2 D_r^2 L^{-2}$; armlength measurement noise is proportional to $D_e^{-1} D_r^{-1} L$ and strain noise proportional to $D_e^{-1} D_r^{-1}$). LISA Pathfinder showed in the frequency range between 0.2 Hz to 5 Hz, the measured displacement noise is (31.9±1.7) fm Hz$^{-1/2}$ agreeing to their OMS model. This is roughly one order above the shot noise and RIN (Relative Intensity Noise). LISA requires about 10 pm Hz$^{-1/2}$ for their arm length measurement at 2 mHz. In the thermally driven OMS Model this is achieved if their Brownian motion is accounted for. In addition, LISA needs to reach this with lower power of incoming light, i.e. shot noise limit should be basically reached. For basic AMIGO, the 12 fm Hz$^{-1/2}$ laser metrology readout noise (i) is already demonstrated within a factor of 3 by LISA Pathfinder in the frequency range between 0.2 Hz to 5 Hz; (ii) is also demonstrated within a factor of 3 by LISA Pathfinder in the frequency range between 0.01 Hz to 0.2 Hz if TTL coupling can be suppressed at this level; (iii) needs demonstration in the 5 Hz to 10 Hz. In addition, in the arm measurement, shot noise needs to be basically reached.

(vii) Many space GW detection proposals need to use constant/equal arm configurations [21]. To name a few, they are AEDGE [9], AIGSO [13, 14]; DECIGO/B-DECIGO [43, 44], ELGAR [18] etc. AIGSO has 10 km arm length, the shortest arm length among these mission proposals. In [21], we calculated that the actuation acceleration needed to maintain such orbits for AIGSO is around 10 pm s$^{-2}$.

In the mission of LISA Pathfinder, different levels of force and torque authority were implemented, from the nominal configuration with x-force authority (on the sensitive line-of-sight axis) of 1100 pm s$^{-2}$ to the URLA configuration levels, with x-force authority of 26 pm s$^{-2}$ [45]. The published LPF differential acceleration noise floor is established by measurements in this configuration. Specifically, LISA Pathfinder demonstrated that when a constant out of the loop force with amplitude of 11.2 pN was applied to the sensitive axis of TM1 (Test Mass 1) for reducing the gravitational imbalance between the TMs, this force does not introduce significant noise or calibration errors [45]. Basically the accelerometer part of the constant-arm technology is already demonstrated by LISA Pathfinder for AIGSO.

B-DECIGO has a nominal arm length of 100 km, DECIGO 1,000 km, and AMIGO 10,000 km. The actuation accelerations needed are respectively 10, 100, and 1000 times more than AIGSO. While the actuation accelerations needed for constant arm implimentation of b-DECIGO and DECIGO is still basically in the LISA Pathfinder nominal configuration range, the actuation accelerations for constant arm AMIGO is one order larger. On what noise level could the actuation accelerations be done needs to be studied and demonstrated carefully for AMIGO. A suggestion is to use an additional test mass (i.e. a pair) to alternate with the original one [7].

(viii) There will be two pathfinder technology demonstration missions with 2 spacecraft/satellites planned in the near future (~ 2025): Taiji-2 and Tianqin-2. Constant arm space interferometry mode could be tested in some stages of the missions (together with the geodetic mode in sequential time frames) if adopted.

(ix) One aim of the mid-frequency GW space missions is to bridge the sensitivity gap between the current/planned Earth-based GW detectors and the mHz space GW detectors under implementation. The optimal arm length would be dependent on the projected sensitivities and the technology that could be achieved at the time of manufacture. So is for AMIGO. In our previous work, we have mentioned that $1 \times 10^7$ m or a few times of this. In the following, we illustrate the noise requirement by a 50,000 km AMIGO termed AMIGO-5:

For the acceleration noise requirement, we set

$S_a^{1/2}(f) \leq 3 \times 10^{-15} [1 + (f/0.3 \text{ Hz})^4]^{1/2}$ m s$^{-2}$ Hz$^{-1/2}$, (10 Hz $> f >$ 10 mHz),     (5)

same as AMIGO before.
For laser metrology noise, we set:

Baseline (b-AMIGO-5): $S_{\text{AMIGOp}}^{1/2} \leq 60$ fm Hz$^{-1/2}$    (10 Hz $> f >$ 10 mHz),     (7)'

Design Goal (AMIGO-5): $S_{\text{AMIGOp}}^{1/2} \leq 19$ fm Hz$^{-1/2}$    (10 Hz $> f >$ 10 mHz),     (8)'

Enhanced Goal (e-AMIGO-5): $S_{\text{AMIGOp}}^{1/2} \leq 2.5$ fm Hz$^{-1/2}$   (10 Hz $> f >$ 10 mHz).   (9)'

Strain ASDs (amplitude spectral densities) vs. frequency for various AMIGO-5 proposals compared to AMIGO proposals are plotted in Fig. 3. The sensitivity curves in the strain power spectral density amplitude vs. frequency plot for AMIGOs of different arm lengths is basically have their flat bottoms shifted to the left in frequency in proportional to the ratio of arm lengths. In our considered frequency range of 10 mHz-10 Hz, the astrophysical confusion limit of LISA/TAIJI/TIANQIN does not play a role.

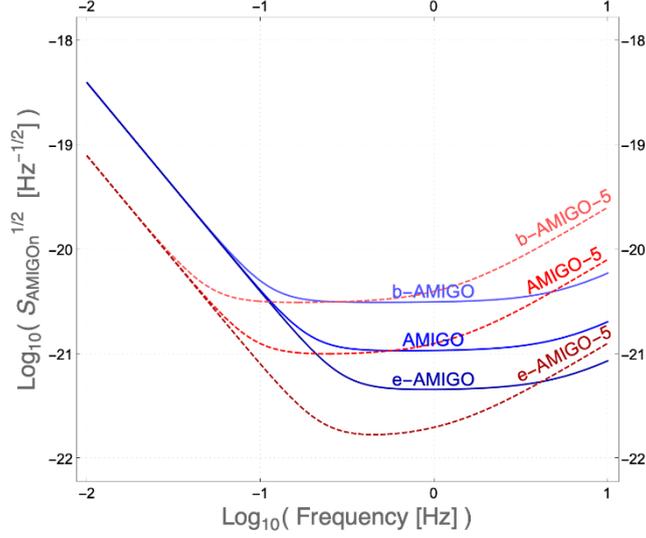

**Fig. 3.** Strain ASDs (amplitude spectral densities) vs. frequency for various AMIGO-5 proposals as compared to AMIGO proposals. The solid lines are for AMIGO's (10,000 km nominal arm length); the dashed lines are for AMIGO-5's (50,000 km nominal arm length)

(x) Taiji-1 [46] and Tianqin-1 [47] have achieved their pathfinder demonstration goals in 2019. After the Taiji-2 and Tianqin-2 pathfinder demonstrations in ~2025 and before the mHz space GW mission launches in 2034, if needed or desired, there could possibly be another pathfinder demonstration or a mid-frequency space GW mission call. Then AMIGO might be a candidate choice for the mission concepts.

The AMIGO-S-8-12deg orbits for 600 days could be an earlier geodetic GW mission option with the orbits worked out starting at a suitable epoch. If a 10-year geodetic mission is desired, it has to go to about 20° behind/leading the Earth orbit. The AMIGO-S-2-6deg orbits for 250 days and the AMIGO-S-2-4deg (for 80 days in the geodetic option; for 300 days or more for the constant equal-arm option) could be a pathfinder mission with one arm (two S/C); they are closer to Earth and takes less days (less than a week to reach the technological demo orbit) and less power for deployments [22].

**Acknowledgement**

I would like to thank Gang Wang and An-Ming Wu for many helpful discussions. I also thank Gang Wang for plotting Fig.1. This work is supported by the Strategic Priority Research Program of the Chinese Academy of Sciences under grant Nos. XDA15020700 and XDB21010100, and by the National Key Research and Development Program of China under Grant Nos. 2016YFA0302002 and 2017YFC0601602.

**References**

1. K. Kuroda, W.-T. Ni and W.-P. Pan, Gravitational waves: Classification, methods of detection, sensitivities, and sources, *Int. J. Mod. Phys. D* **24**, 1530031 (2015); this article is also in One Hundred Years of General Relativity: From Genesis and Empirical Foundations to Gravitational Waves, Cosmology and Quantum Gravity, Chap. 10, ed. W.-T. Ni (World Scientific, Singapore, 2017).
2. I. Mandel, A. Sesana and A. Vecchio, The astrophysical science case for a decihertz gravitational-wave detector, *Class. Quantum Grav.* **35**, 054004 (2018); and references therein.
3. K. A. Kuns, H. Yu, Y. Chen and R. X. Adhikari, Astrophysics and cosmology with a deci-hertz gravitational-wave detector: TianGO, *Phys. Rev. D* **102**, 043001 (2020), arXiv:1908.06004 [gr-qc].
4. M. Arca Sedda, C. P L Berry, K. Jani *et al.*, The Missing Link in Gravitational-Wave Astronomy: Discoveries waiting in the decihertz range, *Class. Quantum Grav.* **37** 215011 (2020), arXiv:1908.11375 [astro-ph.IM]
5. W.-T. Ni, Mid-Frequency Gravitational Wave Detection and Sources, *Int. J. Mod. Phys. D* **29**, 1902005 (2020), arXiv:2004.05590 [gr-qc].
6. B. P. Abbott *et al.* (LIGO Scientific and Virgo Collaborations), GW190521: A Binary Black Hole Merger with a Total Mass of 150 M, *Phys. Rev. Lett.* **125**, 101102 (2020).
7. W.-T. Ni, G. Wang, and A.-M. Wu, Astrodynamical middle-frequency interferometric gravitational wave observatory AMIGO: Mission concept and orbit design, *Int. J. Mod. Phys. D* **29**, (2020) 1940007, arXiv:1909.04995 [gr-qc].
8. A. Sesana, Prospects for multiband gravitational-wave astronomy after GW150914, *Phys. Rev. Lett.* **116**, 231102 (2016).
9. Y. Abou El-Neaj, C. Alpigiani, S. Amairi-Pyka *et al.*, AEDGE: Atomic Experiment for Dark Matter and Gravity Exploration in Space, *EPJ Quantum Technology* volume **7**, 6 (2020), arXiv:1908.00802.
10. J. W. Armstrong *et al.*, *Astrophys. J.* **599**, 806 (2006).
11. J. W. Armstrong, *Living Rev. Rel.* **9**, 1(2006).
12. T. Ebisuzaki, *Int. J. Mod. Phys. D* **28**, 1940002 (2019).
13. D. Gao, J. Wang and M. Zhan, *Commun. Theor. Phys.* **69**, 37 (2018).
14. G. Wang, D. Gao, W.-T. Ni, J. Wang and M.-S. Zhan, *Int. J. Mod. Phys. D* **28**, 1940004 (2019).
15. A. Shoda, Y. Kuwahara, M. Ando, *et al.*, *Phys. Rev. D* **95**, 082004 (2017).
16. T. Shimoda, S.Takano, C.P. Ooi, N. Aritomi, A,Shoda, Y. Michimura, and M. Ando, *Int. J. Mod. Phys. D* **28**, 1940003 (2019), arXiv:1812.01835.
17. B. Canuel et al., *Sci. Rep.* **8**, 14064 (2018).
18. B. Canuel, S. Abend, P. Amaro-Seoane *et al.*, ELGAR—a European Laboratory for Gravitation and Atom-interferometric Research, *Class. Quantum Grav.* **37**, 225017 (2020), arXiv:1911.03701.
19. M.-S. Zhan, J. Wang, W.-T. Ni *et al.*, *Int. J. Mod. Phys. D* **28**, 1940005 (2020), arXiv:1903.09288 [physics.atom-ph].

20. W.-T. Ni, Gravitational Wave (GW) Classification, Space GW Detection Sensitivities and AMIGO (Astrodynamical Middle-frequency Interferometric GW Observatory), Proceedings of Joint Meeting of 13th International Conference on Gravitation, Astrophysics and Cosmology, and 15th Italian-Korean Symposium on Relativistic Astrophysics, Ewha Womans University, Seoul, Korea, July 3-7, 2017, *EPJ Web of Conferences* **168**, 01004 (2018); arXiv:1709.05659 [gr-qc].
21. G. Wang, W.-T. Ni, and A.-M. Wu, Orbit design and thruster requirement for various constant-arm space mission concepts for gravitational-wave observation, *Int. J. Mod. Phys. D* **28**, 1940006 (2020), arXiv:1908.05444 [gr-qc].
22. A.-M. Wu et al., Deployment of AMIGO, paper in preparation.
23. A.-M. Wu, W.-T. Ni and G. Wang, Deployment Simulation for LISA Gravitational Wave Mission, IAC-17-A2.1.4, *68th International Astronautical Congress*, 25-29 September 2017, Adelaide, Australia (2017).
24. A.-M. Wu and W.-T. Ni, Deployment and simulation of the ASTROD-GW formation, *Int. J. Mod. Phys. D* **22**, 1341005 (2013).
25. M. Armano, H. Audley, G. Auger et al., Sub-Femto-g Free Fall for Space-Based Gravitational Wave Observatories: LISA Pathfinder Results, *Phys. Rev. Lett.* **116**, 231101 (2016).
26. P. Amaro-Seoane, H. Audley, S. Babak et al., Laser Interferometer Space Antenna, submitted to ESA on January 13th in response to the call for missions for the L3 slot in the Cosmic Vision Programme, arXiv:1702.00786 [astro-ph.IM].
27. N. Cornish and T. Robson, Galactic binary science with the new LISA design, J. *Phys.: Conf. Ser.* **840,** 012024 (2017).
28. A. Petiteau, M. Hewitson, G. Heinzel, E. Fitzsimons and H. Halloin, LISA noise budget, *Tech. rep. LISA Consortium lISA-CST-TN-0001* (2016).
29. M. Armano, H. Audley and J. Baird *et al.*, Beyond the Required LISA Free-Fall Performance: New LISA Pathfinder Results down to 20 μHz, *Phys. Rev. Lett.* **120,** 061101 (2018).
30. T. Robson, N. Cornish, and C. Liu, The construction and use of LISA sensitivitty curves, *Class. Quantum Grav.* **36** 105011 (2019).
31. J. W. Armstrong, F. B. Estabrook, and M. Tinto, Time-delay interferometry for space-based gravitational wave searches, *Astrophys. J.* **527**, 814 (1999).
32. J. W. Armstrong, F. B. Estabrook, and M. Tinto, Sensitivities of alternate LISA configurations, *Classical Quantum Gravity* **18**, 4059 (2001).
33. G. Wang, W.-T. Ni, W.-B. Han, and C.-F. Qiao, Algorithm for TDI numerical simulation and sensitivity investigation, *Phys. Rev. D*, **103**, 122006 (2021), arXiv:2010.15544.
34. Andrzej Krolak, Massimo Tinto and Michele Vallisneri, Optimal filtering of the LISA data, *Phys. Rev. D* **70** (2004) 022003, *Phys. Rev. D* **76** (2007) 069901 (erratum).
35. Gudrun Wanner, Space-based gravitational wave detection and how LISA Pathfinder successfully paved the way, *Nat. Phys.* **15**, 1 (2019).
36. G. Wanner, Sensor noise in LISA Pathfinder, LISA Symposium, prerecorded, 2020.

37. G. Wanner and N. Karnesis on behalf of the LISA Pathfinder collaboration, Preliminary results on the suppression of sensing cross-talk in LISA Pathfinder, 11th International LISA Symposium, IOP Conf. Series: *Journal of Physics: Conf. Series* **840** (2017) 012043 (IOP Publishing).
38. H. Wegener, V. Müller, G. Heinzel, and M. Misfeldt, Tilt-to-Length Coupling in the GRACE Follow-On Laser Ranging Interferometer, *Journal of Spacecraft and Rockets*, **57**, 1362 (2020).
39. M. Chwalla, K. Danzmann, M. Dovale Álvarez, Optical Suppression of Tilt-to-Length Coupling in the LISA Long-Arm Interferometer, *Physical Review Applied* **14,** 014030 (2020).
40. Tapley, B. D., Bettadpur, S., Watkins, *et al.*, "The Gravity Recovery and Climate Experiment: Mission Overview and Early Results," *Geophysical Research Letters* **31**, 9607 (2004).
41. Kornfeld, R. P., Arnold, B. W., Gross *et al.*, "GRACE-FO: The Gravity Recovery and Climate Experiment Follow-On Mission," *Journal of Spacecraft and Rockets* **56**, 931 (2019).
42. Abich, K., Abramovici, A., Amparan, B., *et al.*, "In-Orbit Performance of the GRACE Follow-On Laser Ranging Interferometer, *Phys. Rev. Lett.* **123**, 031101 (2019).
43. S. Kawamura *et al.*, DECIGO and b-DECIGO, *Int. J. Mod. Phys. D* **27**, 1845001 (2018); and references therein.
44. S. Kawamura, M. Ando, Naoki Seto, et al., Current status of space gravitational wave antenna DECIGO and b-DECIGO, arXiv:2006.13545; and references therein.
45. M. Armano, H. Audley, J. Baird et al., LISA Pathfinder Performance Confirmed in an Open-Loop Configuration: Results from the Free Fall Actuation Mode, *Phys. Rev. Lett.* **123**, 111101 (2019).
46. Xinhua. China Focus: Chinese satellite tests space-based gravitational wave detection technologies, 2019. http://www.xinhuanet.com/english/2019-09/20/c_138408486.htm?from=timeline; https://en.wikipedia.org/wiki/Taiji_Program_in_Space.
47. J. Luo, Y.-Z. Bai, L. Cai, et al., The first round result from the TianQin-1 satellite. *Class. Quantum Grav*. **37**. 185013 (2020); https://en.wikipedia.org/wiki/TianQin.